\documentclass[12pt]{article}

\usepackage{a4wide}
\usepackage{amsmath}
\usepackage{amsfonts}
\usepackage{amssymb}
\usepackage{cite}

\makeatletter

\@addtoreset{equation}{section}
\makeatother

\def\const{(\mathrm{const})}
\def\half{\frac12}
\def\d{\mathrm{d}}

\def \cN{\mathcal N}
\def \Tr{\mathrm{Tr}}
\def \Li{\mathrm{Li}}

\def\mbb{\mathbb }

\def \be { \begin{equation} }
\def \ee { \end{equation} }
\def \beal#1 { \begin{align} #1 \end{align} }
\def \nn {\notag\\}
\def \( {\left( }
\def \) {\right) }
\def \[ {\left[ }
\def \] {\right] }
\def\aver#1{\left\langle {#1} \right\rangle}

\begin{document}

\begin{titlepage}

\title{
\vspace{-2cm}
       \vspace{1.5cm}
       Background Effects on Thermal Free Energy in Conformal Field Theories
       \vspace{1.cm}
}
\author{
Shuichi Yokoyama\thanks{syr18046\textcircled{a}fc.ritsumei.ac.jp}, Ibuki Yoshida\thanks{rp0096hi\textcircled{a}ed.ritsumei.ac.jp}, Hiroki Yunoto\thanks{rp0073eh\textcircled{a}ed.ritsumei.ac.jp}\; 
\\[25pt] 
{\normalsize\it Department of Physical Sciences, College of Science and Engineering,} \\
{\normalsize\it Ritsumeikan University, Shiga 525-8577, Japan}
}

\date{}

\maketitle

\thispagestyle{empty}

\begin{abstract}
\vspace{0.3cm}
\normalsize

We compute background corrections to the thermal free energy of free conformal field theories (CFTs) with a global abelian symmetry defined on a box with all sides infinitely long by turning on a flat gauge field in general dimensions.  
This result is applied to the U($N$) maximally supersymmetric Yang-Mills theory, the ABJ(M) theory, and the free U($N$) vector model in four, three, and two spacetime dimensions, respectively. 
In these examples, the thermal free energies monotonically increase as the background field does, and cross a zero within the valid range. We argue that the systems undergo a phase transition at the zero, where the effective degrees of freedom fall off similarly to the color-confinement one, and that it always occurs if the number of matter fields is greater than a critical value. 
We comment for this phase transition to circumvent the Mermin-Wagner-Coleman theorem in two spacetime dimensions.

\end{abstract}
\end{titlepage}

\section{Introduction} 
\label{Introduction}

When a set of system parameters approaches a critical region, system observables start to exhibit a simple response by scaling laws against variation of system parameters. 
Such a singular behavior of the system at a critical point has attracted a great deal of attention and been intensively studied both experimentally and theoretically \cite{Fisher:1967ttr,RevModPhys.39.395}. 
What has been common to physicists is that the detailed microscopic interactions becomes irrelevant as the system approaches a critical region along renormalization group (RG) flow \cite{PhysRevB.4.3174}, and that the universal behavior of critical phenomena is dictated by conformal symmetry expected to be enhanced at criticality and is classified by conformal field theories (CFTs) \cite{Polyakov:1970xd,Cardy:1984epx}.  

Thermal free energy is a basic observable to see critical phenomena.
For a CFT in $d+1$ spacetime dimensions, thermal free energy density $F$ responds to the variation of the system (absolute) temperature $T$ in such a way that $F/T^{d+1}$ is a constant.
This is a typical example of the insensitivity to the detailed short distance interactions and used to classify the universality of critical phenomena. 
In two spacetime dimensions, this constant is known to be proportional to the conformal anomaly coefficient or the central charge of the Virasoro algebra \cite{PhysRevLett.56.742,PhysRevLett.56.746}, which plays a role to count the degrees of freedom effectively as also implied through RG flow \cite{Zamolodchikov:1986gt}. 
In four dimensions, such a relation to the anomaly coefficients is not clarified yet. Nonetheless, thermal free energy is still a useful observable and its universal constant is expected to count the degrees of freedom contributing to the thermodynamic behavior. This was revealed for example in the context to test the AdS/CFT duality \cite{Maldacena:1997re}, in which the mismatch of the universal constant by a factor of three quarters in a normalization was observed between the expected dual pair of the maximally supersymmetric Yang-Mills theory (SYM) in the weak coupling limit and the maximal supergravity on AdS$_5$ black hole background \cite{Gubser_1996,Gubser_1998}. This mismatch happens because the universal constant depends on the coupling constant in the setup, and the strong/weak duality rather predicts that the universal constant monotonically decreases and asymptotes to three quarters times the value in the weak coupling limit as the coupling constant increases (see also \cite{PhysRevD.59.061701,V_zquez_Mozo_1999}).
The analogous role of thermal free energy was played also in three spacetime dimensions. 
A properly interesting application is to a class of Chern-Simons vector models, which turned out to be a soluble one in the 't Hooft large $N$ limit \cite{Giombi:2011kc,Aharony:2011jz}. In particular, the thermal free energy can be exactly computed in the planar limit \cite{Giombi:2011kc,Jain:2012qi} including chemical potential \cite{Yokoyama:2012fa}, which was used to provide strong evidence for three dimensional Bose-Fermi duality interpolating the level-rank one to the SUSY one known as the Giveon-Kutasov (or Seiberg-like) one \cite{Aharony:2012ns,Jain:2013py,Jain:2013gza,Yokoyama:2013pxa,Moshe:2014bja,Minwalla:2015sca,Geracie:2015drf,Jensen_2019,Minwalla:2023esg}. (See \cite{Wadia:2016zpd} for a review.)

Then it is an interesting problem to both theorists and experimentalists to compute or measure how thermal free energy for an equilibrium system at criticality in heat bath changes according to any change of background, for example, by putting the system into another external reservoir. 
We address this problem by turning on a flat gauge field to a system described by CFT with a global abelian symmetry.  
Turning on background gauge fields for global symmetries and studying their response is a basic tool to analyze the system as seen from its effective use incorporating anomaly coefficients protected from the RG flow \cite{tHooft:1979rat}.
In this paper we are particularly interested in the effect of the constant mode of the background gauge field, which generically gives no effect on a dynamical system on a sufficiently large space. 
However, once the system is placed into heat reservoir and reaches thermal equilibrium, it is described by the same theory formulated on the trivial fibration of a circle over the original space, whose circumference is proportional to the inverse temperature. There arises a gauge invariant operator from gauge field winding on the nontrivial 1-cycle called the Wilson (or 't Hooft) loop one labeled by the zero mode, which gives rise to a nontrivial effect on thermal free energy. A main purpose of this paper is to compute the background corrections to thermal free energy by the gauge zero mode and to apply it for the search of any new physics. 
We organize this paper by starting from general analysis applicable to general dimensions and applying the general result to specific dimensions to see aspects of physics in each dimension. 

\section{General analysis}
\label{generald}

It is convenient to start our study with a free complex scalar field on the $d$-dimensional space $M$ in heat bath at temperature $T$. The thermal free energy for this system can be computed from the partition function of the same theory placed on $M\times\mbb S^1$, where $\mbb S^1$ is the circle of radius $\beta=1/T$ and the scalar field $\Phi$ satisfies the periodic boundary condition for the thermal $\mbb S^1$ direction denoted by $x^D$ with $D=d+1$. 
$
\Phi(\vec x, x^D-\beta) = \Phi(\vec x, x^D). 
$
See \cite{Kapusta:1989tk,lebellac1996thermal,Laine_2016} for textbooks.
If the sides of the space is infinitely long so that $\vec x\in M=\mbb R^d$, the Fourier modes for the scalar field are labeled by continuous momentum variables $\vec p\in \mbb R^d$. 
By performing the gaussian integration for each mode, we obtain the partition function 
\beal{
Z_\Phi =\exp\left[ -V_d \int_{\mbb R^d} \frac{\d^d\vec p}{ (2\pi)^d} \{  \beta \varepsilon  +  2\log (1 - e^{-\beta \varepsilon }) \} \right], 
\label{Z1CS}
} 
where $\varepsilon$ is the energy for one particle state and $V_d$ is the space volume. $V_d$ is so large that we send it to infinity finally. 
We are interested in the critical situation, in which case the one-particle energy is simply given by $\varepsilon=|\vec p|$, so that the above is rewritten as 
\beal{
\lim_{V_d\to\infty}\frac{\log Z_\Phi}{\beta V_d} = - \( \int_{\mbb R^d} \frac{\d^d\vec p}{ (2\pi)^d} |\vec p| \) - \frac1{\beta^{(d+1)}} \int_{\mbb R^d} \frac{\d^d\vec p}{ (2\pi)^d}  2 \log (1 - e^{-|\vec p| }).  
\label{Z1CS2}
} 
The first term is a divergent constant coming from the zero point energy per unit volume and thus a non-universal one dependent on a regularization scheme, while the second term is universal and specified as the minus of thermal free energy.  
In this critical case, the second term can be explicitly evaluated by using the zeta function as 
\beal{
\lim_{V_d\to\infty}\frac{\log Z_\Phi}{\beta V_d} = \const +C_F \frac{\Omega_{d} \Gamma(d) \zeta(d+1) }{ (2\pi)^d \beta^{d+1}}, 
\label{Normalization}
} 
where $\Omega_{d}=\frac{2\pi^{d/2} }{ \Gamma(d/2)}$ is the volume of the unit $(d-1)$-dimensional sphere and $C_F=2$ for the current case of one complex scalar field. $C_F$ counts the degrees of freedom contributing to thermal free energy, and in particular $C_F$ matches the conformal anomaly coefficient or the central charge in two spacetime dimensions turning off the background field \cite{PhysRevLett.56.742,PhysRevLett.56.746}. 
For convenience, we refer to the coefficient $C_F$ as the thermal charge in what follows. 

Now we have fixed the natural normalization, we move on to studying background effects on the thermal charge by turning on flat gauge field for the global $U(1)$ symmetry in the free complex scalar field theory. This is implemented by replacing the partial derivative $\partial_\mu$ into the covariant derivative $D_\mu=\partial_\mu - iA_\mu$. 
As before, we consider a situation where $M=\mbb R^d$.
In this case the flat gauge field in the space directions gives no effect with the spectrum unchanged, while that in the thermal direction must give a nontrivial effect due to the appearance of a gauge invariant Wilson loop winding around the thermal $\mbb S^1$ diretion, which must contribute to thermal free energy. 
This can be easily seen by performing a field redefinition such that $\tilde\Phi(x^\mu)=e^{-ix^\nu A_\nu}\Phi(x^\mu), ~\tilde\Phi^*(x^\mu)=e^{ix^\nu A_\nu}\Phi^*(x^\mu)$. This field redefinition makes the gauge field disappear from the Lagrangian because $(D^\mu \Phi)^* D_\mu \Phi=\partial^\mu \tilde\Phi^* \partial_\mu \tilde\Phi$ and the effect of the flat gauge field is only left so as to change the boundary condition from the periodic one to a twisted one such that 
\be 
\tilde\Phi(\vec x, x^D-\beta) =e^{2\pi i\nu} \tilde\Phi(\vec x, x^D), ~ 
\tilde\Phi^*(\vec x, x^D-\beta) =e^{-2\pi i\nu} \tilde\Phi^*(\vec x, x^D), 
\label{TBC}
\ee
where $\nu=\beta A_D/(2\pi)$ is the exponent of the Wilson loop.\footnote{Note that there is no anomaly in this local U(1) field transformation since the theory is vector-like.}
This implies that the partition function with flat gauge field turned on is expressed in the operator formalism as 
\beal{
\tilde Z_\Phi =\Tr ( e^{2\pi i \nu Q} e^{-\beta H} ), 
\label{ZinOF}
}
where $Q$ is the U(1) charge operator normalized for the field $\Phi$ to have unity and $H$ is the Hamiltonian regarding the thermal $x^D$ direction as the Euclidean time one, and the trace is taken over the same Hilbert space as before. 
In other words, the background effect on the partition function can be included by replacing $\beta\varepsilon$ into $\beta\varepsilon -2\pi i\nu Q$ for the original partition function $Z_\Phi$.
Since the modes from $\Phi$ have $Q=1$ and those from $\Phi^*$ have $Q=-1$, the partition function including the background effect is obtained as  
\be 
\lim_{V_d\to\infty}\frac{\log\tilde Z_\Phi}{V_d} = - \int_{\mbb R^d} \frac{\d^d\vec p}{ (2\pi)^d}  \( \beta |\vec p| + \log (1 - e^{-(\beta |\vec p|+2\pi i\nu) }) +  \log (1 - e^{-(\beta |\vec p| - 2\pi i\nu) }) \) . 
\ee
\footnote{ 
See \cite{Yokoyama:2012fa} for related formulas to direct computation. 
}
Note that the corrections of the gauge field in the non-universal term cancel. 
The second term can be explicitly evaluated by using the polylogarithm, and we reach the following form 
\beal{
\lim_{V_d\to\infty}\frac{\log\tilde Z_\Phi}{\beta V_d} 
=& \const  + \frac{ \Omega_{d} \Gamma(d) }{ (2\pi)^d \beta^{d+1} } (\Li_{d+1}( e^{2\pi i\nu }) +\Li_{d+1}(e^{-2\pi i\nu }) ). 
} 
Therefore the thermal charge including the effect of the flat gauge field to a free complex scalar field theory is determined as
\be 
C_F = \frac{\Li_{d+1}( e^{2\pi i\nu }) +\Li_{d+1}(e^{-2\pi i\nu }) }{\zeta(d+1)}. 
\ee

The thermal charge for a complex fermion field can be similarly computed. 
The number is counted in a way that the Dirac field in $d+1$ spacetime dimensions consists of the $2^{[(d-1)/2]}$ components of complex fermion fields, where $[a]$ stands for the maximum integer which is less than or equal to $a$. 
In particular to four spacetime dimensions, a complex fermion means a Weyl fermion. 
The difference from the bosonic case is that the fermion field $\Psi$ satisfies the anti-periodic boundary condition for the thermal $\mbb S^1$ direction. 
$\Psi(\vec x, x^D-\beta) = -\Psi(\vec x, x^D).$ 
Taking into account the statistics, the thermal charge for a complex fermion is determined as the minus of that of a complex scalar field with $\nu=1/2$: 
$
C_F
= 2(1-2^{-d}). 
$
This result of course reproduces that in \cite{PhysRevLett.56.742,PhysRevLett.56.746} at $d=1$, and that in \cite{Gubser_1996} at $d=3$, and was used for the thermodynamic analysis of $p$-branes \cite{Klebanov_1996}. 

As argued previously, the effect to turn on the flat gauge field reflects as a change of the boundary condition:
\be 
\tilde\Psi(\vec x, x^D-\beta) =-e^{2\pi i\nu} \tilde\Psi(\vec x, x^D), ~ 
\tilde\Psi^*(\vec x, x^D-\beta) =-e^{-2\pi i\nu} \tilde\Psi^*(\vec x, x^D). 
\label{TBCF}
\ee
Then the flat gauge field effect on the thermal charge for a free complex fermion field theory is given by
\be 
C_F =- \frac{\Li_{d+1}(- e^{2\pi i\nu }) +\Li_{d+1}(-e^{-2\pi i\nu }) }{\zeta(d+1)}. 
\ee

As a result, the thermal charge for a free conformal theory of $N_\Phi$ complex scalar fields, $N_\Psi$ Dirac fermions with the flat gauge field turned on the global U(1) symmetry for which the matter fields have unit charge  is determined as  
\be 
C_F =\frac{N_\Phi \{ \Li_{d+1}( \omega) +\Li_{d+1}(\omega^{-1}) \} -2^{[(d-1)/2]} N_\Psi \{ \Li_{d+1}(- \omega) +\Li_{d+1}(-\omega^{-1}) \} }{\zeta(d+1)}. 
\label{MainResult}
\ee
where $\omega= e^{2\pi i\nu }$. 
We shall see applications of this simple result in what follows. 

Comments are in order. 
Firstly, the thermal charge without the effect of the flat gauge field is  
\be 
C_F = 2 \{ N_\Phi  + 2^{[(d-1)/2]} (1-2^{-d}) N_\Psi \}. 
\label{nu0}
\ee
In general space dimension with $d>1$, this is different from the coefficient $C_T$ of the two point function of the symmetric traceless stress tensor in the normalization  
$
\aver{ T_{\mu\nu} (x) T_{\lambda\sigma}(0) }
= \frac{C_T}{\Omega_{d+1}^2} \frac1{(x^2)^{d+1}} {\cal I}_{\mu\nu,\lambda\sigma}(x), 
$
which was determined in \cite{OSBORN1994311} as $C_T=\frac{2(d+1)}{d} N_\Phi +(d+1) 2^{(d-1)/2} N_\Psi $ for the case without the contribution of the vector fields. 

Secondly, the above result can be simplified at even spacetime dimensions as
\be 
C_F =(-)^{\frac{d+1}{2}} (2\pi)^{d+1} \frac{N_\Phi\zeta(-d;\nu)  -2^{(d-1)/2} N_\Psi \zeta(-d;\nu\pm\half)  }{d! \zeta(d+1)}, 
\label{MainResult2}
\ee
by using the Joncqui\`ere's relation obtained by extending the dilogarithm inversion identity 
\cite{Erdelyi1953Higher}, where plus or minus in $\nu\pm\half$ is chosen to be plus for $\nu<\half$ and minus for $\nu>\half$. 
We have checked that this result reproduces that with the gauge field turned off when $\nu=0$ by using a formula of the zeta function $\zeta(1-s) =2^{1-s} \pi^{-s} \cos (\frac{\pi s}2) \Gamma(s)\zeta(s) $. 

Thirdly, the thermal charge and thus the free energy are invariant under the background $\mbb Z_2$ transformation by $\nu\to 1-\nu$. 
This implies that the original theory enjoys a $\mbb Z_2$ symmetry which is broken by turning on the background gauge field. Such a $\mbb Z_2$ action is the Euclidean time refection or the charge conjugation. 
Indeed, the state with charge $Q$ and that with $-Q$ are degenerate in the original $\mbb Z_2$ invariant theory, but they are split after the background gauge field is turned on unless $\nu$ is half. 
This suggests that the broken $\mbb Z_2$ symmetry is restored at $\nu=\half$.\footnote{
Such a symmetry restoration at the middle value of the background field with periodicity was pointed out in a related context \cite{Gaiotto:2017yup}.
}

Fourthly, by performing the analytic continuation for the background gauge field around the origin from a real value to be a pure imaginary one, the partition function obtained above can be regarded as that of the same theory in particle bath whose chemical potential is given by $\nu_0=iA_D$ \cite{Kapusta:1989tk,lebellac1996thermal,Laine_2016}. In our formulation, this is apparent from the expression of the partition function in the operator formalism \eqref{ZinOF}, which is rewritten as  
\beal{
Z =\Tr (  e^{-\beta ( H - \nu_0 Q) } ). 
\label{ZinOF2}
}
This is nothing but the partition function of the grand canonical ensumble. 
The thermal charge for the same theory in the particle bath is given by
\be 
C_F =\frac{N_\Phi \{ \Li_{d+1}(z ) +\Li_{d+1}(z^{-1}) \} -2^{[(d-1)/2]} N_\Psi \{ \Li_{d+1}(- z) +\Li_{d+1}(-z^{-1}) \} }{\zeta(d+1)}
\label{MainResult3}
\ee
where $z= e^{\beta\nu_0 }$. 

Fifthly, if the background gauge field is made dynamical by restricting its form as $\nu=l/p$ with $l=0,1,\cdots, p-1$, the $\mbb Z_p$ subgroup is gauged and an orbifolded CFT is obtained.
The partition function for such an orbifolded CFT is given by summing its twisted component expressed as \eqref{ZinOF} for all possible values of the gauge fields such as $Z=\frac1p\sum_{l=0}^{p-1}\Tr ( e^{2\pi i \frac lp Q} e^{-\beta H} )$, and its thermal charge can be evaluated approximately by extremizing \eqref{MainResult} with respect to $\nu$. 

Finally, the background gauge field gives a decreasing effect on the thermal charge in general.
Therefore, once the background gauge field is turned on, the sign of the thermal charge is not always positive but may become negative as $\nu$ increases. 
Once the sign of the thermal charge flips, the system becomes unstable and undergoes a phase transition in the sense of the breakdown of CFT across the point. Any neglected operator at the conformal fixed point will become relevant and have to be taken into account in order to analyze the theory beyond the transition point. 
Taking into account the negative effect of the background gauge field on the thermal charge in general, it will be concluded that this phase transition always happens if the number of matter fields is greater than a certain critical value.\footnote{ 
The precise condition is obtained from $C_F|_{\nu=1/2}<0$. 
}
We shall see that this type of phase transition happens to all the cases taken below. 

\section{Applications} 
\subsection{To $\cN=4$ U($N$) SYM in $d=3$}
\label{d3}

At $d=3$, the normalization of the thermal charge is \eqref{Normalization}
\beal{
\lim_{V_3\to\infty}\frac{\log Z}{\beta V_3} = \const +\frac{\pi^2}{90} C_F T^4, 
} 
where $C_F$ is given from a formula \eqref{MainResult2} as $
C_F= 240 \{ N_\Phi\zeta(-3;\nu)  - 2 N_\Psi \zeta(-3;\nu\pm\half ) \}.$
\footnote{ 
For convenience, we note the relevant expressions of the Hurwitz zeta function $\zeta(-3;\nu)= -\frac14(\nu^4-2 \nu^3+\nu^2-\frac{1}{30})$, $\zeta(-3;\nu+\half )= -\frac{1}{4}(\nu ^4 -\frac{\nu ^2}{2}+\frac{7}{240})$.
}

Let us apply this result to the $\cN=4$ U($N$) SYM as an example of a four dimensional CFT \cite{Grisaru:1980nk,Avdeev:1980bh,Caswell:1980yi,Sohnius:1981sn,Seiberg:1988ur}. 
As explained in Introduction, the thermal free energy density under the 't Hooft limit is expected to behave as $C_F= 15 N^2  f(\lambda) $, where $f(\lambda)$ describes the dependence of the 't Hooft coupling constant $\lambda$ and was computed as $f(\lambda) = 1 - \frac3{2\pi^2} \lambda+\cdots $ for $\lambda\sim0$ by perturbation of the field theory side \cite{PhysRevD.59.061701}, and $f(\lambda) = \frac34 + \frac{45}{32}\zeta(3) (2\lambda)^{-3/2}+\cdots $ for $\lambda\sim\infty$ by that of the gravity side \cite{Gubser_1998}.

Now let us study the flat gauge background correction to the thermal charge. 
The theory consists of three adjoint chiral multiplets and a vector multiplet, and we turn on the flat gauge field for the baryonic U(1) symmetry. 
By employing the above formula, the thermal charge is computed in the weak coupling limit as $C_F= 15 N^2 \{ 1 - 6\nu ^2 ( 3  - 4 \nu ) \}$ for $\nu<\half$, and $C_F= 15 N^2 \{ 1 - 6(1-\nu)^2 (4 \nu -1) \}$ for $\nu>\half$. 
Note that the non-universal divergent constant cancels for bosons and fermions even with the flat background gauge field turned on. 
Such a cancellation of the non-universal term always happens for any supersymmetric multiplet regardless of the spacetime dimensions. 

One will soon realize that the thermal charge approaches a zero as $\nu$ increases and cross a zero around $\nu\approx 0.3$. Therefore, as argued in the last of the previous section, the $\cN=4$ SYM is expected to undergo a phase transition of the breakdown of conformal symmetric description across the zero point, where the (thermal) degrees of freedom decrease from the order $N^2$ to the order one generically. In this respect, this phase transition is similar to the color confinement one. 

If the thermal charge is denoted by $ C_F= 15 N^2 f(\lambda,\nu) $, this result shows that $f(0,\nu) = 1 - 6\nu ^2 ( 3  - 4 \nu ) $ for $\nu<\half$, and $f(0,\nu)=  1 - 6(1-\nu)^2 (4 \nu -1) $ for $\nu>\half$. 
It would be interesting to compute $f(\lambda,\nu) $ around $\lambda\sim0$ from the field theory side and around $\lambda\sim\infty$ from the gravity side and to investigate how the phase transition point changes as the gauge coupling constant.

\subsection{To the ABJM theory}
\label{d2}

Next we consider the case at $d=2$, in which the normalization of the thermal charge is \eqref{Normalization}
\beal{
\lim_{V_2\to\infty}\frac{\log Z}{\beta V_2} = \const +\frac{\zeta(3)}{2\pi} C_F  T^3, 
} 
where $C_F$ is given from a formula \eqref{MainResult}.

Let us apply this result to a three dimensional CFT of $\cN=6$ ${\rm U}(N)_k \times {\rm U}(M)_{-k}$ Chern-Simons matter theory called the ABJ(M) theory \cite{Aharony:2008ug,Aharony:2008gk}, in which the thermal charge in the case with $M=N$ was computed as $C_F= 7N^2$ in the weak 't Hooft coupling limit $\lambda=N/k\to 0$ and is expected to be $C_F=2^{9/2}\pi^3 N^2/(3^2\zeta(3)\lambda )$ in the strong coupling limit assuming the AdS/CFT duality. 
This implies that the thermal free energy normalized by the thermal charge in the weak 't Hooft coupling limit goes to zero as the 't Hooft coupling constant goes to infinity, which is different from that for $\cN=4$ SYM in four spacetime dimensions. Interestingly, this behavior was observed by explicit computation of the thermal free energy in the 't Hooft limit for a simplest $\cN=2$ U$(N)_k$ Chern-Simons vector model \cite{Jain:2012qi}. 

Now, in order to compute the thermal charge with the flat gauge field taken into account, we recall that the ABJM theory consists of two vector multiplets for U($N$) and U($M$) and four bi-fundamental chiral multiplets.
Since the kinetic term of vector multiplets is the topological one of Chern-Simons interaction, they are expected to have no contribution to the thermal charge in the weak coupling limit. 
Therefore, the thermal charge of the ABJ theory including the flat gauge field is computed as
$C_F= 4NM \{ \Li_{3}( e^{2\pi i\nu }) +\Li_{3}(e^{-2\pi i\nu }) -  \Li_{3}(- e^{2\pi i\nu }) -\Li_{3}(-e^{-2\pi i\nu }) \} / \zeta(3) $. 
This reduces to $7N^2$ in the above case of $M=N, \nu=0$. 
As $\cN=4$ SYM in four spacetime dimensions, the thermal charge for the ABJ(M) theory also has a zero point around $\nu\approx 0.25$, so it is also expected to exhibit a phase transition similar to the color-confinement one due to the effect of the flat gauge field.

\subsection{To the free U($N$) vector model}
\label{d1} 

At $d=1$, the normalization of the thermal charge is 
\beal{
\lim_{V_1\to\infty}\frac{\log Z}{\beta V_1} = \const +\frac{\pi}{6} C_F T^2, 
\label{Normalization1d}
} 
which is fixed to match the conformal anomaly in the case without any background field. 
Turning on the flat gauge field, the thermal charge is corrected as $C_F=- 24 \{N_\Phi\zeta(-1;\nu)  - N_\Psi \zeta(-1;\nu\pm\half ) \}.$ Note that $\zeta(-1;\nu)= -\frac12(\nu^2-\nu+\frac{1}{6})$.

To enhance our understanding of a two dimensional CFT by using this result, let us take a free conformal field theory of $N$ complex scalar fields as an example. 
The thermal or central charge for this system $c$ is $2N$, which is corrected by turning on the flat gauge field. The corrected thermal charge is given by $C_F= 2N(1- 6\nu(1-\nu))$. A question is whether this correction is computable by the two point function of the stress tensor as is for the central charge. 
The answer is no. 
A simple reason is that the local structure of the theory is intact by changing the background so that the operator product expansion of the stress tensor is unchanged. 
Taking into account the invariance of the spectrum, the flat gauge field turned on in the thermal direction plays a role of a grey-body factor. 

This interpretation changes if we compute the partition function in the modular-transformed coordinate system. 
In what follows, we consider the case of $N=1$ just for simplicity. 
Then the theory is placed on a finite circle with circumference $\check L=\beta$ at sufficiently low temperature $\check T=1/\check\beta$ with $\check\beta=V_1$, and the flat gauge field $\check A_{\check\mu}$ is turned on.
To carry out the computation in the operator formalism, we perform the same field redefinition as in section~\ref{generald}.
Then the flat gauge field in the Lagrangian is absorbed into the twisted boundary condition 
\beal{ 
\tilde\Phi(x^{\check1}+\check L, x^{\check 2}) =e^{2\pi i\check\nu_{\check1}} \tilde\Phi(x^{\check 1}, x^{\check 2}), ~ 
\tilde\Phi^*(x^{\check1}+\check L, x^{\check 2}) =e^{-2\pi i\check\nu_{\check1}}  \tilde\Phi^*(x^{\check 1}, x^{\check 2}). \nn
\tilde\Phi(x^{\check1}, x^{\check 2} -\check \beta) =e^{2\pi i\check\nu_{\check2}} \tilde\Phi(x^{\check 1}, x^{\check 2}), ~ 
\tilde\Phi^*(x^{\check1}, x^{\check 2}-\check\beta) =e^{-2\pi i\check\nu_{\check2}}  \tilde\Phi^*(x^{\check 1}, x^{\check 2}). 
\label{TBCx1}
}
where $\check\nu_{\check\mu}=\check L\check A_{\check\mu}/(2\pi)$.
The canonical quantization in this twisted boundary condition was substantially carried out in the context to study an orbifold CFT of a sigma model whose target space is $\mbb C/\mbb Z_p$, in which the conformal dimension of the (twisted) ground state was determined as $h_0=\bar h_0= \half \check\nu_{\check1}(1-\check\nu_{\check1})$ with $0\leq\check\nu_{\check1}<1$ \cite{Dixon:1986qv}. 
This imposes the ground state energy to be given by $E_0= \frac{2\pi}{\check L}(h_0+\bar h_0 - \frac{c}{12})$, where $c=2$. Again, the central charge is unchanged for the same reason stated in the above setup. 
Since the contribution of the ground state becomes dominant in the low temperature limit, $Z \approx e^{-\check \beta E_0}$, we obtain  
\beal{
\lim_{\check\beta\to\infty}(\frac{ \log Z_\Phi }{\check\beta})=- \frac{2\pi}{\check L}(h_0+\bar h_0 - \frac{c}{12}). 
\label{OF1d}
} 
Note that the gauge field contribution in the $x^{\check2}$ direction disappears in the low temperature limit, which is natural with taken into account the disappearance of the (dual) thermal circle. 
By comparing this with \eqref{Normalization1d}, the thermal charge has to be given by $C_F= c - 12 (h_0+\bar h_0 ) $.
This is consistent with the above result by taking account of the modular transformation of the gauge field leading to $\check\nu_{\check1} = 1 - \nu$. 
In this standpoint, the correction to the thermal charge is induced by the energy shift of the ground state, which is nothing but the Casimir energy. 
To see it, let us calculate the partition function in the path-integral formalism. 
By turning on the gauge field in the space $\mbb S^1$ direction, the one-particle energy changes from $|\frac{2\pi n}{\check L}|$ to $|\frac{2\pi (n+\check\nu_{\check1})}{\check L}|$ with $n\in\mbb Z$.
Therefore, the partition function is given by
\beal{
Z_\Phi =\exp\left[ -\sum_{n\in\mbb Z} \{  \check\beta |\frac{2\pi (n+\check\nu_{\check1})}{\check L}|+  \log (1 - e^{-\check\beta |\frac{2\pi (n+\check\nu_{\check1})}{\check L}|+ 2\pi i\check\nu_{\check2}}) +  \log (1 - e^{-\check\beta |\frac{2\pi (n+\check\nu_{\check1})}{\check L}|- 2\pi i\check\nu_{\check2}} ) \} \right]. \notag
}
\footnote{ 
As mentioned in section~\ref{generald}, making the gauge field dynamical by restricting its value in the $\mbb Z_N$ subgroup, the partition function for an orbifold CFT of a sigma model with target space $\mbb C/\mbb Z_p$ is obtained from this one, which is consistent with the one used in a context to analyze fractional D-branes \cite{Bill__2001}. 
}
Again, the contribution of $\check\nu_{\check2}$ cancels in the first term in the exponent. We reach
\beal{
\frac{\log Z_\Phi}{\check\beta} =  -\sum_{n\in\mbb Z}  |\frac{2\pi (n+\check\nu_{\check1})}{\check L}|+ \frac1{ \check\beta} \sum_{n\in\mbb Z} \{\log (1 - e^{-\check\beta |\frac{2\pi (n+\check\nu_{\check1})}{\check L}|+ 2\pi i\check\nu_{\check2}}) +  \log (1 - e^{-\check\beta |\frac{2\pi (n+\check\nu_{\check1})}{\check L}|- 2\pi i\check\nu_{\check2}} )\}. \notag 
}
As previously, the first term is divergent due to the zero point energy and the second one is the finite universal one. However, this does {\it not} mean that the first term is non-universal, because it is now dependent on the background gauge field in the $x^{\check1}$ direction leaving a physical effect known as the Casimir one. 
Extracting it from the divergent term by the zeta regularization, we obtain $-\frac{2\pi}{\check L} (\zeta(-1;\check\nu_{\check1})+\zeta(-1;1-\check\nu_{\check1}) )=\frac{2\pi}{\check L} (\check\nu_{\check1}^2-\check\nu_{\check1}+\frac{1}{6})$.
This exactly matches the result \eqref{OF1d} obtained in the operator formalism by sending $\check\beta$ to infinity, in which the $\check\nu_{\check2}$ contribution has gone away as was discussed. 
It would be amusing that in the original coordinate system the background correction to the thermal charge is obtained from the universal finite part of the free energy by employing the dilogarithm inversion identity, while in the modular-transformed setup the same quantity is computed from the divergent zero point energy by performing the zeta regularization.

\subsection{Comments on the phase transition in two spacetime dimensions} 
\label{Comments}

As (supersymmetric) CFTs in $d=3, 2$ in the previous subsections, the above CFT in $d=1$ is expected to undergo a phase transition because the thermal charge corrected by the background flat gauge field vanishes at $\nu=(3-\sqrt3)/6 \approx 0.2$. 
One might ask whether our argument of the existence of such a phase transition in the two spacetime dimensional CFT contradicts with the Mermin-Wagner-Coleman theorem \cite{PhysRevLett.17.1133,Coleman:1973ci}. 
The answer is no. 
This is simply because the conformal symmetry in the system is broken at the beginning by placing it in heat bath at nonzero temperature with flat gauge field, which is clearly seen from the non-invariance of the partition function and the twisted ground state by the scale transformation or the dilatation, and thus cannot be spontaneously broken at the concerned phase transition point with $C_F=0$, which is a terminal of the CFT description of the two dimensional system. 
This in turn suggests that the Mermin-Wagner-Coleman theorem can be circumvented by breaking such a continuous symmetry that prevents a phase transition by turning on some background field. 
We expect that this phase transition in $d=1$ is similar to those in $d=2, 3$: the system that is well described by the theory of a vector-valued field will be more suitably by that of the singlet one across the zero.

\section*{Acknowledgement}

We would like to thank Yuji Sugawara for valuable discussion and helpful comments.
This work was supported in part by the Grant-in-Aid of the Japanese Ministry of Education, Sciences and Technology, Sports and Culture (MEXT) for Scientific Research (No.~JP26K07091), and by JST SPRING (No.~JPMJSP2101).

\bibliography{TC}

\end{document}